\documentclass[journal,twoside,web]{ieeecolor}
\usepackage{generic}
\usepackage{cite}
\usepackage{amsmath,amssymb,amsfonts}
\usepackage{algorithmic}
\usepackage{graphicx}
\usepackage{algorithm,algorithmic}
\usepackage{eqnarray}
\usepackage{hyperref}
\hypersetup{hidelinks=true}
\usepackage{textcomp}
\def\BibTeX{{\rm B\kern-.05em{\sc i\kern-.025em b}\kern-.08em
    T\kern-.1667em\lower.7ex\hbox{E}\kern-.125emX}}
\makeatletter
\newcommand{\thickhline}{%
    \noalign {\ifnum 0=`}\fi \hrule height 1pt
    \futurelet \reserved@a \@xhline
}
\newcolumntype{"}{@{\hskip\tabcolsep\vrule width 1pt\hskip\tabcolsep}}
\makeatother

\begin{document}

\title{Neuro-TransUNet: Segmentation of stroke lesion in MRI using transformers}
\author{{Muhammad Nouman, Mohamed Mabrok and Essam A. Rashed}
\thanks{This work was supported by JST, PRESTO Grant Number JPMJPR23P7, Japan. The work is also funded by the Qatar-Japan Research Collaboration Research Program under grant number M-QJRC-2023-313. \emph{(Corresponding author: M. Nouman)}}\thanks{M. Nouman and E. A. Rashed is with the Graduate School of Information Science, University of Hyogo, Kobe 650-0047, Japan. M. Mabrok is with the Department of Mathematics, Statistics and Physics, College of Arts and Sciences, Qatar University, Doha, Qatar.}}

\maketitle

\begin{abstract}
Accurate segmentation of the stroke lesions using magnetic resonance imaging (MRI) is associated with difficulties due to the complicated anatomy of the brain and the different properties of the lesions. This study introduces the Neuro-TransUNet framework, which synergizes the U-Net's spatial feature extraction with SwinUNETR's global contextual processing ability, further enhanced by advanced feature fusion and segmentation synthesis techniques. The comprehensive data pre-processing pipeline improves the framework's efficiency, which involves resampling, bias correction, and data standardization, enhancing data quality and consistency. Ablation studies confirm the significant impact of the advanced integration of U-Net with SwinUNETR and data pre-processing pipelines on performance and demonstrate the model's effectiveness. The proposed Neuro-TransUNet model, trained with the ATLAS v2.0 \emph{training} dataset, outperforms existing deep learning algorithms and establishes a new benchmark in stroke lesion segmentation.
\end{abstract}


\section{Introduction}
\label{sec:introduction}
\begin{figure} 
    \centering
    \includegraphics[width=0.45\textwidth]{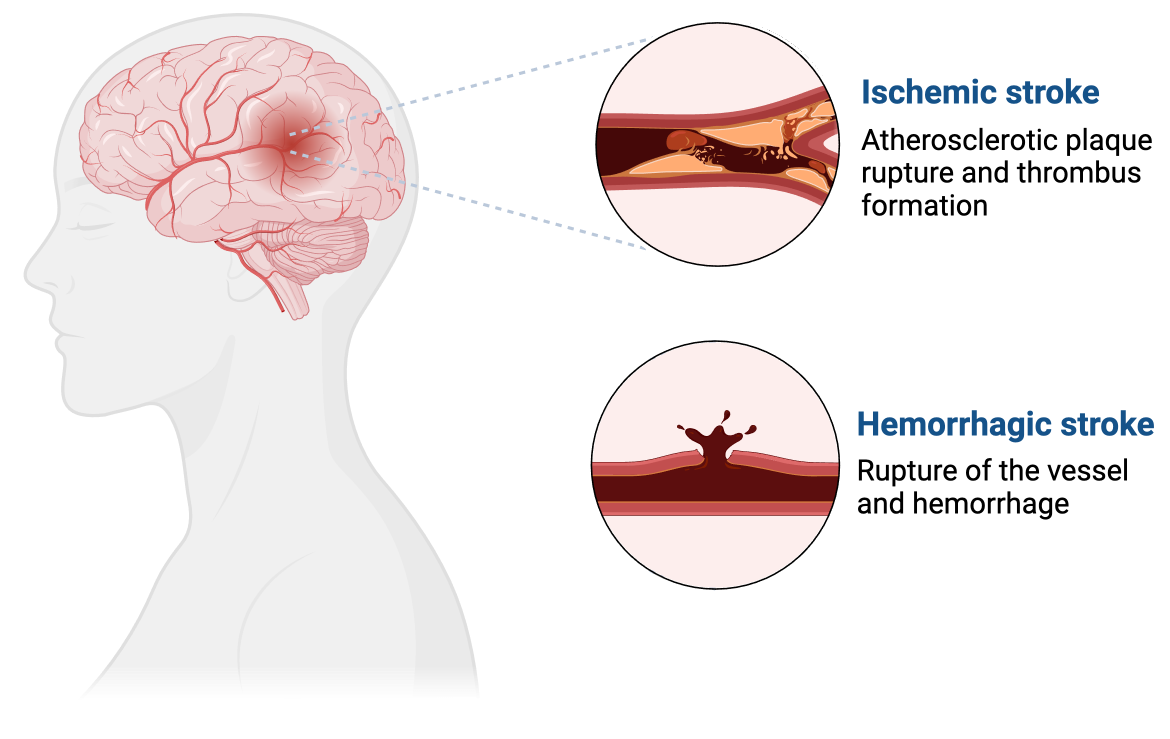}
    \caption{Different types of brain strokes (ischemic and hemorrhagic).}
    \label{stroke}
\end{figure}

\IEEEPARstart{S}{troke} is one of the most formidable challenges to public health across the globe, consistently coming at the top as the leading cause of mortality and long-term disability in diverse populations \cite{who2022stroke,feigin2021global}. According to the World Economic Forum, the incidence of stroke-related deaths is likely to rise significantly, escalating from 6.6 million cases in 2020 to an estimated 9.7 million by 2050 \cite{wef2023strokes}. The growing number of stroke cases underlines the need for innovative technologies, both diagnostic and therapeutic, that can efficiently deal with this problem, by targeting its determinants, symptoms, and effects in terms of disability and death \cite{pu2023projected, feigin2023pragmatic}. Brain stroke is commonly attributed to two primary causes: ischemic stroke, characterized by blockage in blood vessels, and hemorrhagic stroke, resulting from the rupture of vessels leading to bleeding into surrounding tissues (Fig. \ref{stroke}). In neurodiagnostics, the contributions of magnetic resonance imaging (MRI) are irrefutable due to its capability of providing good spatial resolution with which brain structures can be precisely illustrated \cite{jhm2024mri}. Accurate identification and characterization of stroke-affected brain regions is a crucial step towards appropriate treatment and prognosis. The spatial resolution of MRI is commonly 1-2$mm$ for most sequences, which is often enough for most clinical cases \cite{lin2009basic}. The accuracy of MRI is an indispensable part of clinical decision-making, especially in treating acute ischemic stroke \cite{yuen2022portable,zhao2021deep}. In addition, the multi-spectral nature of MRI makes it invaluable in the post-diagnosis stage, where it helps in understanding the viability of the affected brain tissues \cite{patil2022detection}.

The past decade has been associated with major milestones in medical image analysis, attributed to integrating sophisticated computational technologies \cite{liu2021advances}. Deep learning methods, particularly convolutional neural networks, have guided a new era of automatic medical imaging, revolutionizing conventional segmentation challenges \cite{shen2017deep,chen2022recent}. Models such as SwinUNETR are the visualization of this transition. They are fast, accurate, and scalable solutions that reduce the variability engendered by human interpretation \cite{he2023swinunetr,cao2022swin}. SwinUNETR, an innovative transformer-based model, has reached the highest performance level on different medical image segmentation tasks; among these are the BTCV Multi-organ Segmentation Challenge and the Medical Segmentation Decathlon (MSD) \cite{tang2022self}. For instance, such technological advancements have proven their importance, even in resource-constrained settings where radiological expertise is limited~\cite{najjar2023redefining}.

\begin{figure} 
    \centering
    \includegraphics[width=0.48\textwidth]{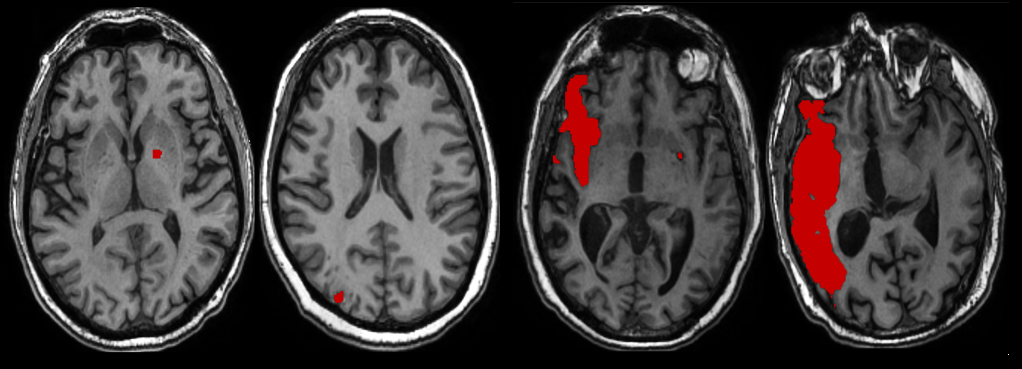}
    \caption{Diverse representations of stroke (in red) in MRI scans.}
    \label{sample}
\end{figure}

Identifying and segmenting stroke lesions from MRI is a critical aspect of selecting the right clinical interventions and formulate an accurate forecast about the patient’s recovery~\cite{tomita2020automatic}. Figure \ref{sample} shows diverse types of strokes in different sizes and locations, making it difficult to accurately identify and delineate these pathologic phenomena. However, despite these technological advancements, current models still have challenges in accurately detecting stroke lesions of varying sizes within unified diagnostic framework \cite{liu2023deep}. 

Another problem that has been well-known for a long time is the data imbalance, caused by small lesions being underrepresented compared to big ones \cite{malik2024stroke}. Although transformers have performed very well at capturing global contextual information, they usually fail to perform localized, meticulous tasks. Transformers use the self-attention mechanism that focuses on long-range dependencies and global features, with a tendency to overlook fine-grained details \cite{chen2021transunet}. Common deep learning models perform well at recognizing either small, medium, or large stroke lesions, but they can hardly detect all sizes concurrently \cite{subudhi2022application}. This highlights a crucial gap: the need for models that have the flexibility to detect stroke lesions of various sizes simultaneously and with high accuracy. This work expands on the preliminary findings presented in \cite{Nouman2024} by extending the analysis and incorporating additional results.

This paper aims to fill the gap in current research by proposing a novel deep learning framework that is developed to address the challenge of segmenting ischemic stroke in T1 3D MRI. Ischemic strokes occur more frequently than other kinds of strokes, which makes accurate and timely identification crucial. By ingeniously integrating the SwinUNETR architecture \cite{hatamizadeh2021swin} with the U-Net framework \cite{soleimani2023utilizing} through advanced feature fusion and segmentation synthesis. The proposed framework (called, Neuro-TransUNet) connects the strengths of both. This defined synthesis allows the acquisition of comprehensive analysis, which provides a high level of accuracy and sensitivity for lesion localization. The contributions of this study are: a) develop a deep learning model that improves the precision of stroke lesions segmentation in 3D MRI, b) attempt to solve the data imbalance problem by creating a comprehensive pre-processing pipeline that also improves the model ability on different lesion sizes and shapes, c) validate the Neuro-TransUNet effectiveness on the challenging ATLAS v2.0 benchmark dataset. 

This paper is organized as follows: related work investigates the current state of segmentation methodologies and deep learning for medical images. The methodology describes the preprocessing pipeline along the Neuro-TransUNet architecture. Training protocols and the experiment results show the model implementation details and results, followed by a discussion and conclusion with future directions.

\section{Related Work}

Recently, there have been notable developments in stroke lesion segmentation using 3D MRI. The ATLAS v2.0 dataset \cite{liew2022large} is a key tool for the evaluation and benchmarking of new methodologies. Table \ref{related1} summarizes previous studies, including the aim, dataset insights, preprocessing methods, structure, architecture descriptions, and loss functions. This compilation is not only an exhibit of the variety and complexity of modern approaches but it also a contributor to the deepening of knowledge regarding this subject. The studies are stacked together to form the foundation of the model’s methodology. Following the technical explanation in Table \ref{related1}, there is a need for further elaboration on the major findings and methodological innovations in these studies. It provokes an in-depth study, and Table \ref{related2} is supplemented with more detailed information on the innovations and deficiencies of these recent studies. This comprehensive summary serves as the basis for the proposed model. Rather than highlighting only the achievements and breakthroughs of previous studies, methodology is designed to overcome the current limitations. 

\begin{table*}
\caption{Overview of stroke lesion segmentation research utilizing ATLAS v2.0.} 
\begin{tabular}{|p{3cm}|p{1.5cm}|p{2.5cm}|p{3cm}|p{2cm}|p{2cm}|p{0.5cm}|}
\thickhline
{\bf Purpose} & {\bf Dataset} & {\bf Preprocessing} & {\bf Methods} & {\bf Structure} & {\bf Loss function} & {\bf Ref.} \\
\thickhline
Improve the accuracy of stroke lesion segmentation in MRI using a novel FISRG algorithm & ATLAS v2.0 & Gaussian denoising & Fuzzy Information
Seeded Region Growing (FISRG) with k-means clustering for seed selection and mathematical morphology for post-processing & Combines fuzzy logic with SRG techniques & Not specified & \cite{gonzalez2023fuzzy}\\
\hline
Segment small-size stroke lesions from MRIs using HCSNet & ATLAS v2.0&2D image slicing, matrix complement, clipping&Hybrid Contextual Semantic Network (HCSNet) with an encoder-decoder architecture and HCSM&U-shaped architecture with HCSM&Mixing-loss function combining dice loss and focal loss&\cite{liu2023hybrid}\\
\hline
Joint learning for stroke lesion segmentation and TICI grading using SQMLP-net &ATLAS v2.0&Intensity correction, MNI-152 template registration, and normalization&SQMLP-net with a segmentation branch and a classification branch&Hybrid multi-task network with shared encoder&Joint loss function combining segmentation and classification losses&\cite{liu2023simulated}\\
\hline
Benchmark various deep supervised models for stroke lesion segmentation & ATLAS v2.0&Z-score normalization, slice handling for 2D models&Evaluation of deep supervised U-Net style models on 2D and 3D MRI&Multiple U-Net variants (traditional, residual, and attention-based)&Not specified&\cite{deb2023best}\\
\hline
Enhance brain lesion segmentation using large-kernel attention within a U-Net architecture & ATLAS~v2.0, ISLES, and BraTS&ATLAS: skull-stripping, bias-correction, re-slicing; ISLES: re-slicing, augmentation; BraTS: re-slicing, skull-stripping, cropping&Convolutional transformer block variant in U-Net architecture with large-kernel attention and post-processing&U-Net with transformer blocks&Dice and cross-entropy loss&\cite{chalcroft2023large}\\
\hline
Improve stroke lesion segmentation accuracy through adaptive image harmonization & ATLAS v2.0&Normalization, data augmentation, and perturbation techniques&Adaptive region harmonization (ARH) for foreground and background alignment&ARHNet architecture&Reconstruction, boundary-aware total variation, and adversarial loss&\cite{huo2023arhnet}\\
\hline
Meta-analysis of transfer learning for brain lesion segmentation & ATLAS~v2.0, In-house dataset&Resampling, skull stripping, image slicing, and data augmentation&Mixed data approach and intermediate task training (ImTT) using transfer learning&Various 2D deep learning architectures&Not specified&\cite{mohapatra2023meta}\\
\hline
Improve the segmentation of biomedical images by addressing instances of imbalance & ATLAS v2.0&Intensity normalization, data augmentation (flipping, rotation, zooming, shifting)&Combined instance-wise and center-of-instance loss with comparisons to dice loss and blob loss&3D Residual U-Net&ICI loss combining global dice, instance-wise, and center-of-instance losses&\cite{rachmadi2024improving}\\
\hline
Improve the generalization ability of stroke lesion segmentation using deep learning & ATLAS v2.0&Registered to the MNI-152 template, normalization&nnU-Net based model with various training schemes, ensemble techniques, and post-processing&nnU-Net, an
encoder-decoder architecture&Default compound loss (dice plus cross-entropy)&\cite{huo2022mapping}\\
\hline
Improve the chronic stroke lesion segmentation & ATLAS v2.0&Intensity normalization, MNI-152 template registration, HD-BET, and data augmentation&Deep Neural Network (DNN) using 3D-UNet with 5-fold cross- validation&3D-UNet for volumetric segmentation&Not specified&\cite{verma2022automatic}\\
\thickhline
\end{tabular}
\label{related1}
\end{table*}

\begin{table*} 
\caption{Contribution, highlights, and identified research gaps of related work listed in Table~\ref{related1}.}
\begin{tabular}{|p{5cm}|p{5cm}|p{5cm}|p{0.5cm}|}
\thickhline
{\bf Contribution} & {\bf Highlights} & {\bf Gap} & {\bf Ref.} \\
\thickhline
Developed the FISRG algorithm which improves stroke lesion segmentation using fuzzy logic with seeded region growing techniques. & Demonstrated high accuracy in processing various appearances of lesions on MRI scans; offers a promising basis for stroke diagnosis. & The algorithm needs to be improved in terms of its capability to discriminate lesions from neighboring areas that have the same intensity and adaptability to the topological changes. & \cite{gonzalez2023fuzzy}\\
\hline
Proposed the HCSNet, combining spatial and channel contextual features and the design of the encoder-decoder architecture, for a precise segmentation of the small-sized stroke lesions. & The network had demonstrated superior performance in segmenting small lesions, benefiting from the fused contextual features capability. & HCSNet needs to be evaluated in more detail concerning its generalization ability for different lesion sizes and types. & \cite{liu2023hybrid} \\
\hline
Introduced the novel method of simultaneous segmentation and grading while leveraging quantum mechanics simulations for learning. & Demonstrated well-performed feature sharing and task optimization, which enabled its multi-task learning performance. & It is necessary to find the right balance between the trade-offs of multi-task learning weights more efficiently. & \cite{liu2023simulated} \\
\hline
Conducted a thorough comparison of the different U-Net architecture designs for both 2D and 3D MRI of stroke lesion segmentation. & Pointed out that the U-Net architecture can tackle large datasets and be used in both 2D and 3D modalities. & It indicates the need for a deeper investigation of transformer models in 3D segmentation and the use of the model on large datasets. & \cite{deb2023best} \\
\hline
Proposed a U-Net model with large kernel attention methods to take advantage of the strengths of CNN and transformer to perform 3D brain lesion segmentation. & The model showed competitive performance with the use of an efficient number of parameters and a special focus on long-range spatial dependencies. & There is a need for research on different kernel sizes as well as embeddings of the patches. & \cite{chalcroft2023large} \\
\hline
Introduced the novel ARHNet model that targets the issue of image intensity disparity by integrating intensity perturbation and region-adaptive harmonization. & Demonstrated the superiority of the proposed method in image harmonization to ensure the strength of augmented images. & Differentiating small lesions remains a challenge; understanding the features of different lesions is still a problem. & \cite{huo2023arhnet} \\
\hline
Implemented transfer learning and mixed data techniques across 2D deep learning models to improve stroke lesion segmentation accuracy. & The application of transfer learning and ensemble methods increased the accuracy, especially by using the novel agreement window technique. & Further investigation is needed on scalability to 3D models and more detailed performance metrics for specific models. & \cite{mohapatra2023meta}\\
\hline
Introduced instance-wise and center-of-instance (ICI) loss to deal with instance imbalance in biomedical image segmentation using a 3D residual U-Net architecture. & The demonstrated ICI loss certainly improved performance, especially in segmenting small instances and addressing imbalances between instances. & More validation of ICI loss's effectiveness is needed for a larger number of segmentation tasks and datasets. & \cite{rachmadi2024improving} \\
\hline
Developed a segmentation framework with nnU-Net, including ensemble learning and post-processing techniques to improve the model generalization. & Demonstrated top performance in the MICCAI ATLAS Challenge, which evaluated the efficacy against various lesion sizes and shapes. & The problem remains for tiny or infrequent spots; future improvements are needed to optimize segmentation accuracy. & \cite{huo2022mapping} \\
\hline
Implemented an automated segmentation model based on the 3D-UNet architecture for chronic stroke lesion detection. & Obtained a proficient level of segmentation accuracy, creating a base for automated evaluation of chronic stroke lesions. & There is a need for research in the field of training dataset bias and for independent evaluation to confirm the model's generalizability capability. & \cite{verma2022automatic} \\
\thickhline
\end{tabular}
\label{related2}
\end{table*}

\section{Methodology}

The Neuro-TransUNet framework is designed to enhance the accuracy of lesion segmentation in MRI. This section explains the proposed framework's dataset description, preprocessing pipeline, and network architecture.

\subsection{Dataset}
\begin{figure} 
\centering
\includegraphics[width=0.45\textwidth]{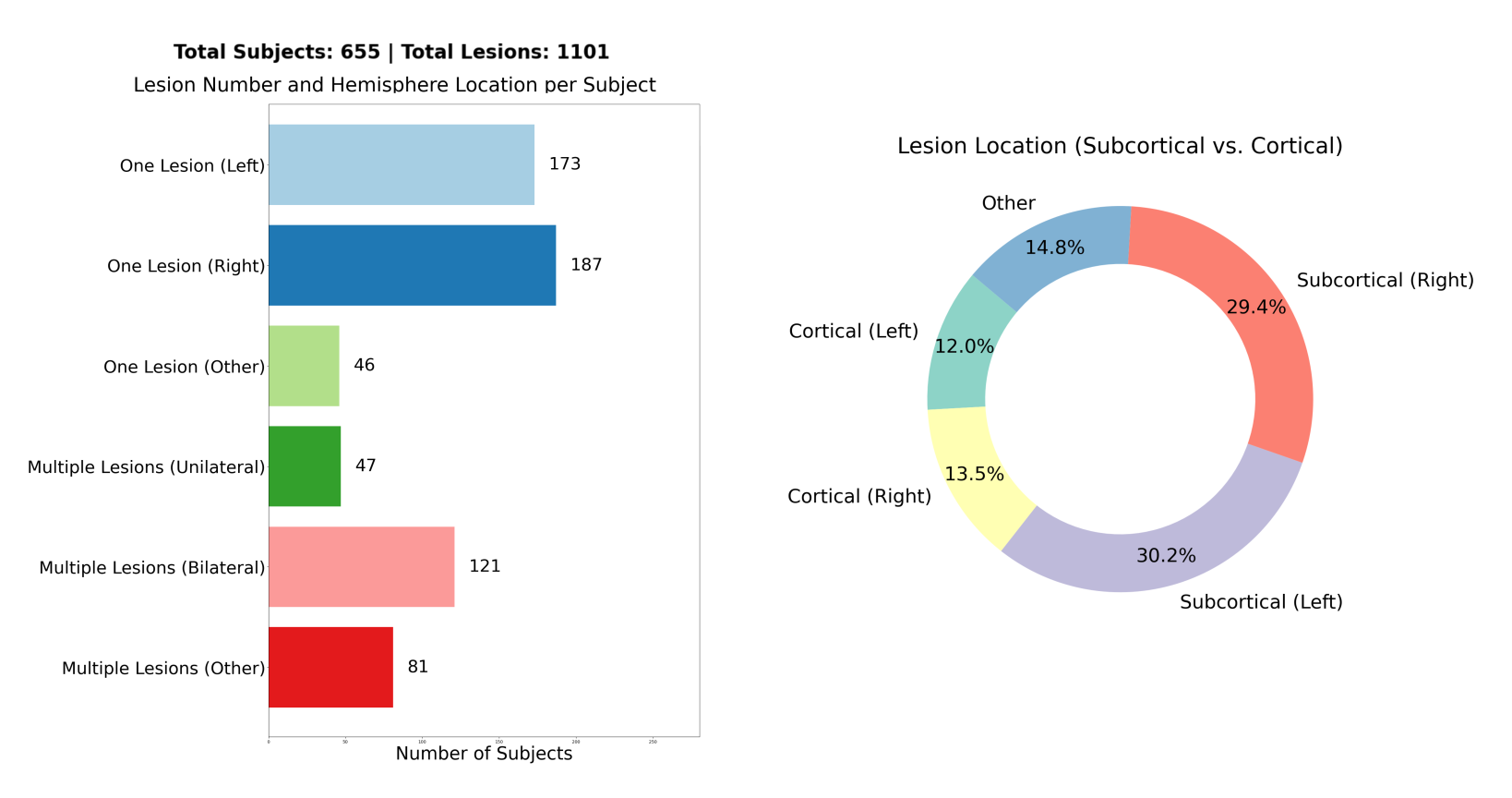}
\caption{Stroke distribution in the ATLAS v2.0 training dataset. The bar plot shows the number of subjects with either single or multiple lesions, subdivided into specific hemisphere locations. The pie chart shows the percentage of lesions identified in different regions.}
\label{lesionplot}
\end{figure}

The ATLAS v2.0, a well-curated dataset of 3D T1-weighted MRI subjects~\cite{liew2022large}. The dataset is the extended version of ATLAS v1.2, and it comprises 1271 MRI subjects with manually labeled lesion masks. The dataset is divided into three subsets: 1) Training set: comprising 655 MRI subjects with ground truth values; 2) Testing set: which includes 300 MRIs with hidden lesion masks; and 3) Generalization set: contains 316 MRIs with data not previously exposed to during the training phase, testing the model's ability to generalize. The dataset is diverse and spans forty-four cohorts and eleven countries, providing a robust representation of the various types of lesions, their locations, and sizes. The diversity will result in the creation of a model that can be broadly applied to a variety of clinical conditions. The dataset incorporates a range of lesion sizes, locations, and appearances and compiled to cover a wide spectrum of lesion features, including single and multiple lesions in different cerebral hemispheres (Fig.~\ref{lesionplot}). 


Another challenge is brain anatomy's complexity and the multispectral features of strokes. The complexity is a step higher in 3D images, where the edges of the lesions from the surrounding brain tissue become particularly important in differentiation. This causes difficulty in identifying lesions accurately and segmenting them using simple methods. Also, manual lesion segmentation is a subjective process that may lead to a bias in the training dataset. This study employs the ATLAS v2.0 training dataset, which includes 655 MRI subjects. Due to the unavailability of ground truth values in the testing subset, the training is split (80\%, 20\%) for training and testing, respectively.

\subsection{Data pre-processing pipeline}
The model's accuracy is highly dependent on a data pre-processing infrastructure that is systematically designed. The pipeline applies a series of operations to the raw MRI to make it suitable for the advanced deep learning model. This pipeline unifies the N4Bias field correction and adaptive resampling methods to ensure the input data is high-reliability and consistent.

\subsubsection{Resampling}
\sloppy
It began by loading an MRI dataset with the help of the \texttt{load\_nifti} function in the NIfTI format to preserve data integrity and consistency. One of the core steps here is substituting zeros for nan (not a number) values to ensure smooth model training. This is accomplished through the \texttt{replace\_nans\_with\_zero} function, which examines each image for NaN values and replaces them with zeros. This method guarantees data consistency and prevents computational errors that could disrupt subsequent analyses. Then, the subjects are resampled to standardized voxel dimensions, a vital step to ensure compatibility with different scanning protocols and equipment. This standardization is done by using custom-made functions, which are used to guarantee unambiguousness across the dataset. Two interpolation techniques are employed for resampling: 'nearest' interpolation for binary masks that helps preserve lesions boundaries, and 'linear' interpolation for image data that promotes continuity of intensity values. This process aims to preserve the accuracy of the spatial model of the brain, which means that for all the subjects, spatial dimensions are uniform.


\subsubsection{Bias correction}
We employ the N4Bias Field Correction algorithm \cite{tustison2010n4itk}, a technique focusing on intensity variation correction across images. Magnetically inhomogeneous intensities of magnetic fields can also be seen in MRI due to different scanner calibrations and magnetic fields. These heterogeneities may not provide clear details of the images and make it difficult to differentiate pathological features. The N4Bias field correction algorithm uses a non-linear filtering method. The function \texttt{bias\_field\_correction} utilizes SimpleITK's \texttt{N4BiasFieldCorrectionImageFilter}, a non-linear image filtering process. This process contributes to reducing the impact of external variability in data. The corrected pictures display improved clarity and uniformity.


\subsubsection{Data standardization and augmentation}
This last step of the pre-processing pipeline highlights the necessity of image normalization and a data augmentation strategy. The normalization of MRI to the range of [0, 1] is accomplished by the scaling intensity values. The scanning is performed on image-to-image to ensure that there is a single, unified dataset. This scaling not only makes uniform input distribution simpler but it also improves the model's ability to detect subtle differences in tissue characteristics.  \normalfont Normalization is vital to make the model work with data having the same statistical properties, which are heavily based on gradient-based optimization techniques. This step also includes error handling to ensure robustness and reliability, particularly when image retrieval or processing fails. Following normalization, image resizing is conducted to a consistent size of 160×160×160. This standardization of all images implies that the images have a uniform size. A standardized input size facilitates the model's learning process because of the uniformity of the input size. Moreover, the data augmentation technique formulated by the MONAI framework is employed to synthetically enlarge the dataset by simulating natural variations. Methods like random flipping, rotation \cite{cirillo2021best}, and affine transformations can simulate the disorder observed in real-world anatomical structures. Each transformation is carefully designed to allow the model to generalize from training data to previously unseen clinical images.


\subsection{Neuro-TransUNet network architecture}
\begin{figure*}
\centering
\includegraphics[width=0.95\textwidth]{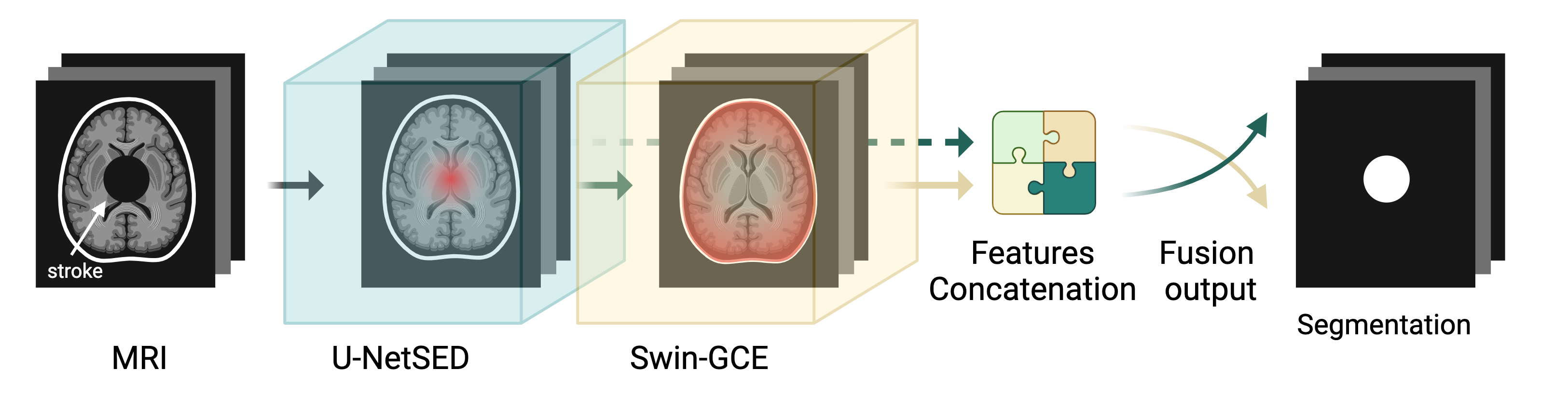}
\caption{A semantic visualization of the Neuro-TransUNet architecture, emphasizing the synergistic combination of the spatial and contextual information processing components. It effectively illustrates the model’s three pivotal components: the U-Net spatial encoder-decoder (U-NetSED), the SwinUNETR global context encoder (Swin-GCE), and the feature fusion and segmentation synthesis mechanism.}
\label{newmodel}
\end{figure*}

The proposed Neuro-TransUNet architecture sequentially integrates the functionality of U-Net with the adaptability of SwinUNETR to effectively manage the challenges of MRI. This integrated procedure combines the local features extracted by U-Net with the global contextual information provided by SwinUNETR. This architecture (Fig.~\ref{newmodel}) is designed to inspect brain structures and provide detailed segmentation of stroke lesions.

Neuro-TransUNet, unlike the classical methods that apply these models separately, integrates these models into a single comprehensive scheme. The Swin-GCE includes an attention mechanism, making it suited for the high-dimensional data typical of MRI. The neural network architecture becomes more sophisticated by adding an extra depth of layers and employing dropout and batch normalization methods at selected points. It follows a structured pipeline of processing that not only preserves but also improves each model’s strengths. These adaptations ensure that the transformer's power in capturing extensive contextual information complements U-Net's precision in local detail extraction.

\subsubsection{U-Net spatial encoder-decoder (U-NetSED)}

The U-Net structure is symmetric with the encoder path that holds context and the decoder path that gives high-precision localization. The encoder accomplishes this by using convolution layers to sequentially downsample the input image, extracting features while retaining relevant contextual information at multiple scales. The deep convolutional layers in the architecture have variable kernel sizes, which are intended to optimize the network to meet the different shapes and sizes of stroke lesions.
\begin{equation}
f_i^n(x) = \text{PReLU}\left(BN\left(Conv_n\left(f_i^{n-1}(x) \right)\right) + x\right).
\end{equation}

At the $i$-th layer, the convolution operation is denoted as Conv$_i$ ($x$) ranging from 1 to n. Following that, batch normalization is used to normalize each convolution's output. Finally, PReLU is applied to each normalized output after each convolution. To deal with variability in stroke lesions, the U-Net part has been modified by increasing the channel depth and adding some targeted residual connections to optimize its capability. These are designed to enhance the detection and characterization of stroke lesions. Every layer incorporates dropout and batch normalization, which helps the learning process and prevents overfitting. The increased number of channels at each convolutional block allows the architecture to detect a wider range of features, making it sensitive to different lesion characteristics. Residual connections are added to provide a stable gradient flow during the training. This helps to stabilize the training process and allows more complex patterns to be learned without the gradient vanishing. The output feature maps of U-Net become the input for the SwinUNETR, bringing the local spatial features of U-NetSED into the contextual knowledge of Swin-GCE.

\subsubsection{SwinUNETR global context encoder (Swin-GCE)}
The Swin-GCE component is designed to capture global contextual information. The SwinUNETR architecture takes advantage of the Swin Transformer architecture and complements it with shifted windowing schemes for self-attention. This adaptation helps to keep computational costs down and ensures that dependencies can be captured over long ranges. Unlike traditional transformers that address self-attention through the entire input space, SwinUNETR focuses on self-attention within the local windows. A technological component is to shift the window from layer to layer, ensuring comprehensive coverage and integration of global context throughout the imaging volume. The SwinUNETR's operation in the framework is expressed as follows:

\begin{equation}
\mathcal{F}_\text{Swin}(x) = x + \text{MLP}(LN(x + SW-MSA(LN(x)))).
\label{eq2}
\end{equation}

Here, the input $x$ denotes the output of the feature map from U-NetSED, which is subsequently transferred to SwinUNETR. The shifted window multi-head self-attention ($SW-MSA$) output is added to the input of the original $x$, forming a residual connection. The sum is then normalized using layer normalization ($LN$). Then the $MLP$ block receives the normalized feature set and provides the model with better presentable features. To solve the vanishing gradient problem, a residual connection is added to the $MLP$ output, where the original input $x$ is taken as the addition. SwinUNETR integration following U-NetSED feature map extraction is essential, as U-Net's detailed local features would enable the SwinUNETR to focus on the global context for accurate segmentation. The high processing capacities of SwinUNETR, in conjunction with the refined details of the enhanced feature maps from U-Net, improve the model's performance.

\subsubsection{Features fusion and segmentation synthesis}
The last part of the architecture is the feature fusion mechanism, which is a fusion of U-NetSED with Swin-GCE. The fusion process starts by concatenating the feature maps from the U-NetSED and Swin-GCE:
\begin{equation}
\mathcal{F}_\text{combined} = \text{Concat}(\mathcal{F}_\text{U-NetSED}, \mathcal{F}_\text{Swin-GCE}),
\label{eq3}
\end{equation}
where $F_\text{U-NetSED}$ represents the local detailed feature map and $F_\text{Swin-GCE}$ is the globally contextualized features. Following concatenation, the combined features are processed through a fusion convolution layer.
\begin{equation}
\mathcal{F}_\text{fused} = \text{Conv3d}(\mathcal{F}_\text{combined}).
\label{eq4}
\end{equation}
This convolutional layer performs the 1$\times$1$\times$1 kernel operation to blend and compress the channel dimensions. The processed features undergo batch normalization and ReLU activation for a nonlinear feature transformation:
\begin{equation}
\mathcal{F}_\text{activated} = \text{ReLU}(BN(\mathcal{F}_\text{fused})).
\label{eq5}
\end{equation}
To ensure the model's robustness and prevent overfitting, dropout is applied to  $\mathcal{F}_\text{activated}$:
\begin{equation}
\mathcal{F}_\text{regularized} = \text{Dropout}(\mathcal{F}_\text{activated}).
\label{eq6}
\end{equation}
The final segmentation output is refined through additional convolutional layers (kernels=3$\times$3$\times$3 and 1$\times$1$\times$1) to map the processed features to a final output.
\begin{equation}
\mathcal{F}_\text{out} =  \text{Conv3d}(\text{Conv3d}(\mathcal{F}_\text{regularized})).
\label{eq7}
\end{equation}

The layers of the network that come after the fusion further optimize the combined features, changing the dimensionality of the features while preserving spatial details. It results in a very detailed segmentation map, $\mathcal{F}_\text{out}$. The entire feature fusion and segmentation synthesis process can be formalized as a sequence of transformation functions applied sequentially to the input features:
\begin{equation}
\mathcal{F}_\text{out} = \mathcal{T}_\text{N}(\dots(\mathcal{T}_\text{2}(\mathcal{T}_\text{1}(\mathcal{F}_\text{in})))).
\label{eq9}
\end{equation}

\sloppy
Individually, all the transform stages $\mathcal{T}_i$ are carefully produced. They all include convolution, batch normalization, activation, and dropout operations. The entire process of feature fusion, as well as the segmentation synthesis, reflects the Neuro-TransUNet architecture's innovative approach.

\section{Implementation details and training protocol}

\subsection{Experiment setup}
The model is executed in a Jupyter Notebook environment that allows the necessary dynamic interaction at the development and testing levels. Transparency, reproducibility, and the ability to easily repeat or expand the work without deviation are the primary forces driving the experiment. Table \ref{tab4} shows the hardware and software used in the experiment. 

\begin{table}
\centering
\caption{Configuration Details of the Experimental Environment}
\begin{tabular}{|l|l|}
\thickhline
{\bf Component} & {\bf Specification} \\
\thickhline
GPU & Nvidia RTX 6000 Ada \\
\hline
OS & Ubuntu 22.04 \\
\hline
Memory & 128 GB\\
\hline
Software & Cuda 12.2 \\
& Python 3.11.5 \\
& PyTorch 2.1.2 \\
& MONAI 1.2.0 \\
& NumPy 1.23.5 \\
\thickhline
\end{tabular}
\label{tab4}
\end{table}

\subsection{Inference process}
The proposed model takes MRI as full 3D volumes to capture the full spatial context of the data representation. This volumetric approach extends MRI analysis beyond the traditional slice-by-slice analysis. The model implements 3D volumes of MRI to obtain high-resolution and detailed information about the lesion's volume, shape, and spatial correlation. That information is not so clear in a slice-based analysis. Volumetric analysis is a key component that is critical for making clinical decisions, such as determining the influence of lesions on overall brain function. By using images in 3D, the model avoids the shared challenges that come with slice-based methods, which are used to separate individual slices and can lead to segmentation errors due to the slice-by-slice treatment.

\section{Experiments and results}

\begin{figure*}
\centering
\includegraphics[width=0.9\textwidth]{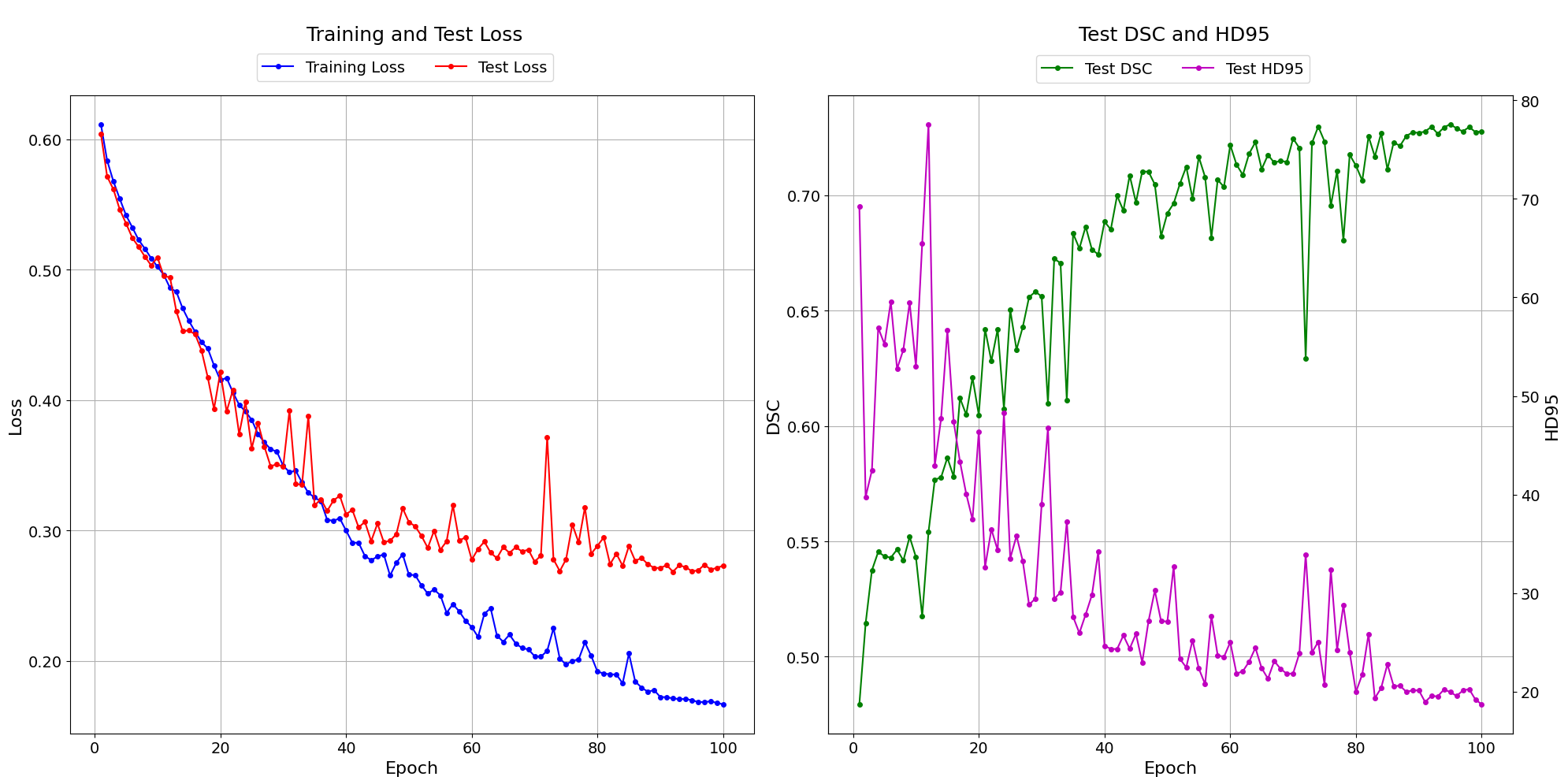}
\caption{The model performance across 100 epochs. The training loss that is gradually reduced shows the model is learning, while the test loss trend is also reducing, indicating that the model has generalization capability. The DSC demonstrates increasing accuracy in segmenting the desired regions. The Test HD95 shows a downward trend, indicating better precision in boundary segmentation}
\label{performance}
\end{figure*}

The proposed model is a sequential integration of the spatial precision of U-Net and the contextual depth of SwinUNETR. The evolution of the dice score along with test loss and hausdorff distance, provides an objective insight into the model’s capability to manage complex segmentation tasks (Fig.~\ref{performance}). A thorough assessment included comparing the model's results with other algorithms. When evaluated on the ATLAS v2.0 dataset, the Neuro-TransUNet yielded optimal results: a test dice score of 0.730 with an average training time of 698.39 seconds per epoch. The model showed a stable increase in the dice score to precisely segregate stroke lesions, with an overall gain of 52.49\% from its initial iteration. An improvement in the dice score shows a positive relationship with correct lesion identification that can promote effective patient care.

The HD95 fluctuated but demonstrated a decreasing tendency, which amounts to 72.99\%, and an optimal value of 18.70. This metric is of the utmost importance for applications that need precise boundary detection and shows that the model has enhanced precision in the contouring of lesions. The ASSD also improved, decreasing by 96.30\% to a final value of 1.269. It indicates the model's enhanced precision in boundary segmentation. The comparative analysis of the model with state-of-the-art (SOTA) is crucial for the positioning of Neuro-TransUNet. Table~\ref{diceval} shows the dice scores of each model categorized by their score nature, which shows the proposed model performance with these methodologies. Compared to the previous models, the performance of the Neuro-TransUNet is highlighted.

\begin{table*} 
\centering
\caption{Comparative analysis of SOTA on ATLAS v2.0.}
\begin{tabular}{|p{3cm}|c|c|p{11cm}|c|}
\thickhline
{\bf Model} & {\bf DSC} & {\bf Scoring} & {\bf Performance analysis} & {\bf Ref.} \\
\thickhline
X-Net & 0.313 & 2D & Moderate complexity, prone to overfitting & \cite{woo2023comparison} \\
\hline
UNETR & 0.347 & 3D & High parameters, lower efficiency in 3D tasks & \cite{woo2023comparison} \\
\hline
SwinUnet & 0.448 & 2D & Efficient feature integration, complexity in training & \cite{woo2023comparison} \\
\hline
Residual U-Net & 0.504 & 3D & Requires balancing between depth and computational load & \cite{deb2023best} \\
\hline
3D-ResU-Net & 0.512 & 3D & Computationally intensive for 3D data & \cite{woo2023comparison} \\
\hline
SegNet & 0.533 & 2D & Balances efficiency and performance, less detailed & \cite{woo2023comparison} \\
\hline
PSPNet & 0.580 & 2D & Requires significant computational resources & \cite{woo2023comparison} \\
\hline
Residual U-Net (ICI loss)& 0.581 & 3D & High computation due to complex instance-wise loss functions & \cite{rachmadi2024improving} \\
\hline
U-net Transformer & 0.583 & 2D & Transformer integration increases complexity & \cite{deb2023best} \\
\hline
HarDNet & 0.591 & 2D & Optimized for speed, lack feature extraction & \cite{woo2023comparison} \\
\hline
U-Net & 0.598 & 2D & Limited by simplicity in handling complex patterns & \cite{woo2023comparison} \\
\hline
Ensemble (PP) & 0.667 & 3D$^*$ & Complex ensemble, intensive post-processing & \cite{huo2022mapping} \\
\hline
LKA-ED & 0.678 & 3D$^*$ & Optimization needed for varying kernel sizes & \cite{chalcroft2023large} \\
\hline
LKA-E & 0.682 & 3D$^*$ & Balances efficiency with performance, slightly better from LKA-ED & \cite{chalcroft2023large} \\
\hline
HCSNet & 0.697 & 3D$^*$ & Specialized for small lesion detection, high complexity & \cite{liu2023hybrid} \\
\hline
SQMLP-net & 0.709 & 3D$^*$ & Multi-tasking increases model complexity and training time & \cite{liu2023simulated} \\
\hline
{\bf Neuro-TransUNet} & {\bf 0.730} & 3D & Advanced architecture requires systematic integration & \\
\thickhline
\end{tabular}
\raggedright \emph{$^*$Average DSC scores.}
\label{diceval}
\end{table*}

Neuro-TransUNet achieves better segmentation accuracy by sequentially integrating deep learning structures without using any pre-trained models or post-processing. Table~\ref{diceval} shows that even though the Neuro-TransUNet, despite being a 3D model, achieves better results in 3D, it also has a better dice score than the highest-performing 2D model. The analysis covers various techniques, including 2D models like SwinUnet and PSPNet, that often operate faster; however, they lack the depth perception required for lesion segmentation. Like UNETR and 3D-ResU-Net, which are sophisticated 3D models and use volumetric data to provide the overall anatomical structures, crucial for precise segmentation.

\begin{figure*}
\centering
\includegraphics[width=0.9\textwidth]{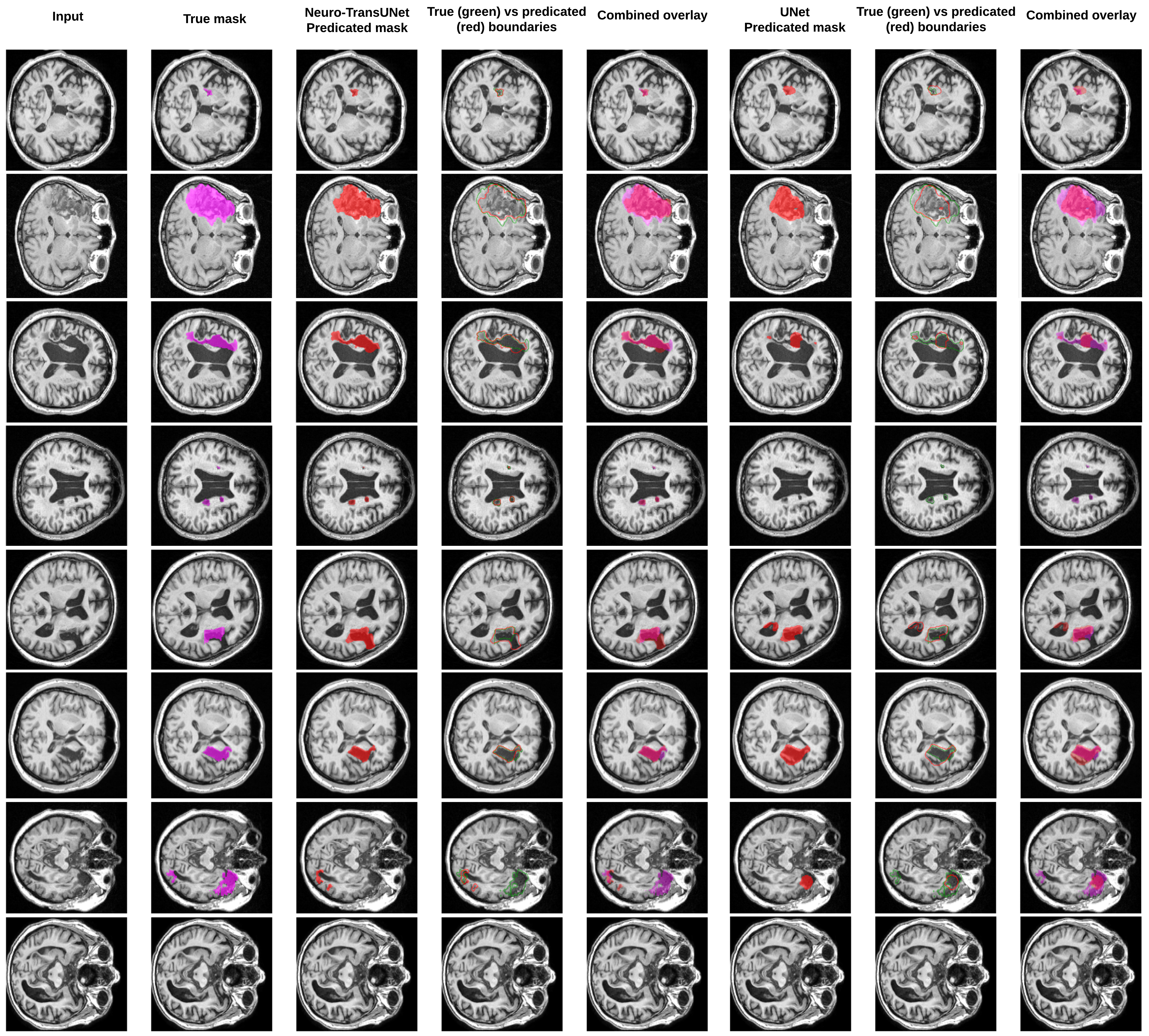}
\caption{Visualization depicts a systematic evaluation of Neuro-TransUNet’s performance on MRI. Each row has a different patient case and a flow that starts with the MRI, goes to the ground truth, the predicated mask overlay by Neuro-TransUNet, a comparison of true versus predicted boundaries, and ends with an overlay of all the elements. The following right side shows the model-predicated masks without SwinUNETR, a comparison of predicted boundaries, and a combined overlay. It also emphasizes the improvements made by incorporating SwinUNETR into the Neuro-TransUNet model. The visualizations display a range of lesion presentations, ranging from small to numerous lesions, including subjects with no lesions to those with multiple lesion occurrences.}
\label{comparison}
\end{figure*}

Models that use large-kernel attention approaches, such as LKA-E and LKA-ED, can strike a balance between computational efficiency and the ability to produce accurate segmentation results. One of the 3D residual U-Net variants that use instance-wise and center-of-instance segmentation loss functions aims to improve the model's performance, considering the size disparities between the ischemic stroke lesions. The FISRG model,  based on the fuzzy information seeded region growing method, has been shown to yield a high dice score of 0.942. Nevertheless, this is achieved with the use of single patient data. These factors, including detailed post-processing, even for one patient's data, led to the fact that FISRG is not included in the comparative study. This contributes to the growing understanding that such models need to be able to generalize across different clinical scenarios. The proposed model's test segmentation results for the stroke lesions are depicted in Fig.~\ref{comparison}.

The precision of the Neuro-TransUNet model in segmenting small, large, and multiple lesions is good, as it has shown a close match with the annotations of experts. Such correspondence suggests that the model can detect clinically different changes in the neuropathological state. The model has demonstrated the ability to precisely locate small lesions, which are more difficult to detect than larger lesions due to their low contrast and signal-to-noise ratio.  The model's specificity stands out in lesion-free subjects, a particularly essential factor used in the prevention of false positives and in avoiding the risk of wrong diagnosis. Furthermore, the model can correctly identify multiple lesions that might be present in a single case, which is necessary for a complete evaluation of the stroke lesions.
The visualization results also reveal the effectiveness of integrating SwinUNETR into the proposed model. The SwinUNETR, which provides global context, significantly enhances the segmentation of lesions. In cases with multiple lesions, the proposed Neuro-TransUNet model also yields better accuracy in segmenting the lesions. This improved performance highlights the importance of incorporating SwinUNETR into more advanced fusion and segmentation synergistic approaches for better lesion identification in complex medical images. In the quantitative metrics, U-Net shows the optimal dice score of 0.633, a 13.29\% decrease compared to the proposed model. The HD95 also decreased from 18.70 to 23.74, demonstrating the efficacy of incorporating SwinUNETR to improve model performance. 
The proposed model tends to have technical limitations when delineating lesions with non-distinct edges or non-standard diffusion patterns, where the segmentation may unconsciously underestimate the actual size of the lesion. In these situations, the segmentation scenarios may deviate from the reality of the task of distinguishing pathological tissue from the brain's natural textural background. Such deviations, on the other hand, point out that the algorithmic interpretation of the same brain lesions is closely tied to their complex morphological profiles, indicating room for model improvement. The emergence of the above complexity implies a demand for larger data sets to train with more paths of pathological variants. The inclusion of more data with multiple modalities and potentially unusual appearances can improve the model's learning process. Sophisticated training would help Neuro-TransUNet become acquainted with the unpredictable variability in medical imaging, allowing it to generalize more effectively. This will bring about more precise segmentation outputs that can precisely depict the complexities of cerebral abnormalities.

\subsection{Impact of preprocessing pipeline}
This section analyzes the effect of two preprocessing pipelines on the segmentation of stroke lesions using a Neuro-TransUNet framework. The comparison evaluates the model's performance using a comprehensive and basic preprocessing approach. This study utilized fifty-two subjects to ensure a thorough assessment, which comprised a wide variation of stroke lesion characteristics. This smaller dataset was chosen to guarantee a representation of different lesion types and dimensions, accompanied by various pathologic homogeneities.

The comprehensive preprocessing included a well-defined protocol that comprised resampling, bias field correction, skull stripping, resizing, intensity normalization, and data augmentation and standardization. The model yielded a dice score of 0.74 with the comprehensive pipeline. However, this study revealed that the skull stripping process introduced noteworthy mistakes in accurately determining the cerebral limits on the 3D MRI. The distinguished detail in the brain shape in the 3D brain mapping algorithms made it difficult to differentiate the brain tissue from other non-brain structures accurately. Uniformly applying a set of all parameters for skull stripping across all subjects could lead to the unintended removal of important brain structures. To ensure accuracy, each subject must be processed individually for skull stripping, which is a time-consuming and challenging process. Thus, it is deemed prudent to exclude skull stripping from the comprehensive preprocessing scheme for the full training dataset, despite its inclusion in this subset. The basic preprocessing involves only the essential processes, with resampling, bias adjustment, and skull stripping purposefully left out. The model achieved the best dice score of 0.65 using the basic preprocessed data, indicating that the model's accuracy was reduced without comprehensive preprocessing.

\subsection{Model parameters}
The model design is expanded by a detailed analysis of the parameters, which shows the vast learning capacity that contributes to its high segmentation accuracy. The network architecture with 100,076,263 trainable parameters alone is a full-fledged testimony to how detailed and complex the architecture can be to achieve results. The distribution of model parameters is also considered, with careful layer allocation allowing for highly accurate feature extraction and representation. The initial layers, such as unet.model.0.conv.unit0.conv, contain parameters in the hundreds, setting the stage for the model's feature learning. As the depth increases, so does the parameter count, culminating in layers with parameters in the millions.



\section{Discussion and conclusion}
The main purpose of this research is to conduct a detailed analysis of the Neuro-TransUNet architecture in terms of its ability to segment stroke lesions from 3D MR images with high precision. The research focused on assessing the architecture and pre-processing strategies used to improve the model's segmentation accuracy. The outcome of this study shows that Neuro-TransUNet attained a test dice score of 0.730 using a volumetric approach that bypasses traditional post-processing, 2D, and 3D models. The proposed model performs well in detecting small, medium, large, and multiple lesions in a subject by sequentially integrating the U-Net and SwinUNETR using advanced feature fusion and segmentation synthesis techniques. The U-Net spatial features improved the SwinUNETR's global contextual ability to detect large and multiple lesions in a subject. This approach reflects the natural anatomical complexity of the brain, making the model accurate and viable for stroke lesion segmentation over a wide range of appearances. This execution displays the Neuro-TransUNet's effectiveness in comparison to recent techniques used in similar tasks.

The findings provide insight into various data processing pipelines and advanced neural network designs for accuracy in clinical diagnostics. These not only extend the current understanding of neural network applications in medical imaging but also underline the importance of these factors. The research found specific situations where the model's segmentation precision is still lacking, showing that more data, can be added to the training dataset to boost the model's ability to analyze complex medical images. The model's results are satisfactory, but the study has some shortcomings that need to be explored further.  The model can conduct the segmentation task well, but sometimes it might have some challenges, such as the diversity and complexity of stroke lesions. Training the model with larger, more diverse datasets can improve its robustness and generalization ability. In addition, the feasibility of working with a system as computationally complex as this model in a typical clinical environment needs to be considered. 

Future research will focus on implementing advanced techniques such as meta-learning or few-shot learning, which can improve the model's adaptability and generalizability. The notion of assessing the applicability of multimodal data, such as functional MRI or diffusion tensor imaging, in addition to imaging, might result in a more elaborate and precise delineation of the lesion. Furthermore, improving interpretable methodologies to understand the model’s decision-making process could reveal the driving mechanisms of segmentation performance.  Based on the data, the design imperfections can be identified and used for the next generation of architecture and to assist the clinician in interpreting the results.

\bibliographystyle{IEEEtran}
\bibliography{IEEEabrv,Refs}

\end{document}